\documentclass[12pt]{nature}

\spacing{1}
\spacing{2}
\spacing{1}

\usepackage{graphicx}

\newcommand{\DP}{\mbox{$\Delta P$}}
\newcommand{\DPobs}{\mbox{$\Delta P_{\rm obs}$}}
\newcommand{\DPg}{\mbox{$\Delta P_{\rm g}$}}
\newcommand{\Dnu}{\mbox{$\Delta \nu$}}
\newcommand{\kepler}{\mbox{\textit{Kepler}}}

\renewcommand{\kepler}{\mbox{{Kepler}}}

\newcommand{\muHz}{\mbox{$\mu$Hz}}
\newcommand{\numax}{\mbox{$\nu_{\rm max}$}}

\newcommand{\paul}{6928997}

\newcommand{\new}[1]{\textbf{\em #1}}
\renewcommand{\new}[1]{{#1}}

\newcommand{\apj}{Astrophys. J.}
\newcommand{\apjl}{Astrophys. J.}
\newcommand{\apjs}{Astrophys. J. Suppl. Ser.}
\newcommand{\aap}{Astron. Astrophys.}
\newcommand{\mnras}{Mon. Not. R. Astron. Soc.}

\newcommand{\nat}{Nature}
\newcommand{\apss}{Astrophys. Space Sci.}

\let\citep\cite
\let\citet\cite

\title{Gravity modes as a way to distinguish between hydrogen- and helium-burning
red giant stars}

\begin{document}

\author{
Timothy R.~Bedding$^{1}$, 
Benoit Mosser$^{2}$, 
Daniel Huber$^{1}$, 
Josefina Montalb\'an$^{3}$, 
Paul Beck$^{4}$, 
J{\o}rgen Christensen-Dalsgaard$^{5}$, 
Yvonne P.~Elsworth$^{6}$, 
Rafael A.~Garc\'\i a$^{7}$, 
Andrea Miglio$^{3,6}$, 
Dennis Stello$^{1}$, 
Timothy R.~White$^{1}$, 
Joris De~Ridder$^{4}$, 
Saskia Hekker$^{6,8}$, 
Conny Aerts$^{4,9}$, 
Caroline Barban$^{2}$, 
Kevin Belkacem$^{10}$, 
Anne-Marie Broomhall$^{6}$, 
Timothy M.~Brown$^{11}$, 
Derek L.~Buzasi$^{12}$, 
Fabien Carrier$^{4}$, 
William J.~Chaplin$^{6}$, 
Maria Pia Di~Mauro$^{13}$, 
Marc-Antoine Dupret$^{3}$, 
S{\o}ren Frandsen$^{5}$, 
Ronald L.~Gilliland$^{14}$, 
Marie-Jo Goupil$^{2}$, 
Jon M.~Jenkins$^{15}$, 
Thomas Kallinger$^{16}$, 
Steven Kawaler$^{17}$, 
Hans Kjeldsen$^{5}$, 
Savita Mathur$^{18}$, 
Arlette Noels$^{3}$, 
Victor Silva~Aguirre$^{19}$ \& 
Paolo Ventura$^{20}$ 
}
\maketitle
\begin{affiliations}
\small
\item Sydney Institute for Astronomy (SIfA), School of Physics, University of Sydney, NSW 2006, Australia
\item LESIA, CNRS, Universit\'e Pierre et Marie Curie, Universit\'e Denis Diderot, Observatoire de Paris, 92195 Meudon cedex, France
\item Institut d'Astrophysique et de G\'{e}ophysique de l'Universit\'{e} de Li\`{e}ge, All\'ee du 6 Ao\^{u}t 17 - B 4000 Li\`{e}ge, Belgium
\item Instituut voor Sterrenkunde, K.U.Leuven, Celestijnenlaan 200D, 3001 Leuven, Belgium
\item Danish AsteroSeismology Centre (DASC), Department of Physics and Astronomy, Aarhus University, DK-8000 Aarhus C, Denmark
\item School of Physics and Astronomy, University of Birmingham, Birmingham B15 2TT, UK
\item Laboratoire AIM, CEA/DSM-CNRS, Universit\'e Paris 7 Diderot, IRFU/SAp, Centre de Saclay, 91191, Gif-sur-Yvette, France
\item Astronomical Institute `Anton Pannekoek', University of Amsterdam, Science Park 904, 1098 XH Amsterdam, The Netherlands
\item IMAPP, Department of Astrophysics, Radboud University Nijmegen, PO Box 9010, 6500 GL Nijmegen, The Netherlands
\item Institut d'Astrophysique Spatiale, UMR 8617, Universit\'e Paris XI, B\^atiment 121, 91405 Orsay Cedex, France
\item Las Cumbres Observatory Global Telescope, Goleta, CA 93117, USA
\item Eureka Scientific, 2452 Delmer Street Suite 100, Oakland, CA 94602-3017, USA
\item INAF - IASF, Istituto di Astrofisica Spaziale e Fisica Cosmica, Via del Fosso del Cavaliere 100, 00133 Roma, Italy
\item Space Telescope Science Institute, 3700 San Martin Drive, Baltimore, Maryland 21218, USA
\item SETI Institute/NASA Ames Research Center, MS 244-30, Moffett Field, CA 94035, USA
\item Department of Physics and Astronomy, University of British Columbia, 6224 Agricultural Road, Vancouver, BC V6T 1Z1, Canada
\item Department of Physics and Astronomy, Iowa State University, Ames, IA 50011, USA
\item High Altitude Observatory, NCAR, P.O. Box 3000, Boulder, CO 80307, USA
\item Max-Planck-Institut f\"ur Astrophysik, Karl-Schwarzschild-Str.~1, 85748 Garching, Germany
\item INAF - Osservatorio Astronomico di Roma, Via Frascati 33, 00040 Monte Porzio Catone (RM), Italy
\end{affiliations}

\clearpage

\begin{abstract}
Red giants are evolved stars that have exhausted the supply of hydrogen in
their cores and instead burn hydrogen in a surrounding
shell\citep{S+H62,Ibe68}.  Once a red giant is sufficiently evolved, the
helium in the core also undergoes fusion\citep{S+G78}.  Outstanding issues
in our understanding of red giants include uncertainties in the amount of
mass lost at the surface before helium ignition and the amount of internal
mixing from rotation and other processes\citep{Cha2005}.  Progress is
hampered by our inability to distinguish between red giants burning helium
in the core and those still only burning hydrogen in a shell.
Asteroseismology offers a way forward, being a powerful tool for probing
the internal structures of stars using their natural oscillation
frequencies\cite{AChDK2010}.  Here we report observations of gravity-mode
period spacings in red giants\cite{BBM2011} that permit a distinction
between evolutionary stages to be made.  We use high-precision photometry
obtained with the {\em Kepler} spacecraft over more than a year to measure
oscillations in several hundred red giants.  We find many stars whose
dipole modes show sequences with approximately regular period spacings.
These stars fall into two clear groups, allowing us to distinguish
unambiguously between hydrogen-shell-burning stars (period spacing mostly
$\sim$50 seconds) and those that are also burning helium (period spacing
$\sim$100 to 300 seconds).
\end{abstract}

Oscillations in red giants, like those in the Sun, are thought to be
excited by near-surface convection.  The observed oscillation spectra are
indeed remarkably Sun-like, with a broad range of radial and non-radial
modes in a characteristic comb
pattern\citep{DeRBB2009,KWB2010,BHS2010,HBS2010,MBG2011}.  However,
theoretical models of red
giants\citep{DGH2001,ChD2004,DBS2009,MMN2010,DiMCC2011} reveal a more
complicated story for the non-radial modes (those with angular degree~$l\ge
1$), \new{and it has been suggested that this offers a means to
determine the evolutionary states of these stars\citep{MMN2010}.}  Owing
to the large density gradient outside the helium core, a red giant is
effectively divided into two cavities.  In the envelope, the oscillations
have properties of acoustic pressure modes (p~modes), but in the core, they
behave like gravity modes (g~modes), with buoyancy as the restoring force.
The models predict a very dense spectrum of these so-called mixed modes for
each value of~$l$ (except $l=0$, as radial g~modes do not exist).  Most
mixed modes have a much larger amplitude in the core than in the envelope
and we refer to them as g-dominated mixed modes.  Like pure g~modes, they
are approximately equally spaced in period\citep{Tas80,MME2008} and
measuring their average period spacing (\DP) would give a valuable new
asteroseismic probe of the cores of red giants.  Unfortunately, they have
very high inertias (the total interior mass that is affected by the
oscillation), which leads to a very low amplitude at the stellar surface
and makes them essentially impossible to observe.  However, because of
resonant coupling between the two cavities, some of the mixed modes have an
enhanced amplitude in the envelope, making them more like p~modes.  These
\new{p-dominated} mixed modes have a lower inertia than the g-dominated
mixed modes, and so their amplitudes can be high enough to render them
observable.  We expect their frequencies to be shifted from the regular
asymptotic spacing, a feature known as `mode bumping'\cite{ASW77}.

\begin{figure}
  \begin{center}
    \includegraphics[width=0.9\textwidth]{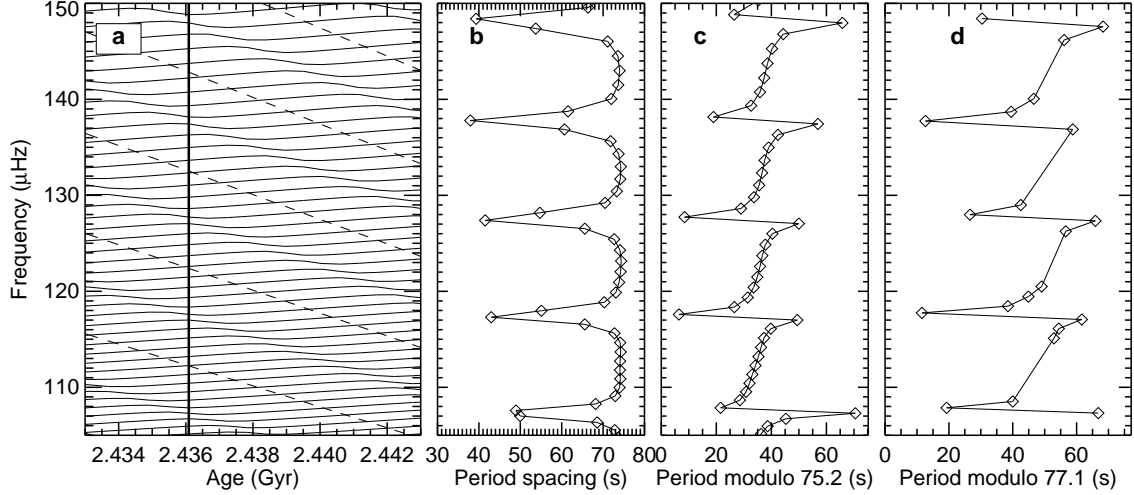}
  \end{center}

  \caption{{\sffamily\textbf{Mixed modes and avoided crossings in red giant stars.}}
    {\sffamily\textbf a,}~Evolution with time of oscillation frequencies in
    a model of a hydrogen-shell-burning red giant with a mass of
    1.5$M_\odot$ and solar metallicity, calculated using the Aarhus stellar
    evolution code \textsc{astec}\cite{ChD2008a}.  Dashed lines show radial
    modes ($l=0$) and solid lines show dipole modes ($l=1$).  The models
    span ranges in radius and luminosity of 6.3 to 6.7\,$R_\odot$ and 19.1
    to 21.4\,$L_\odot$, respectively (here $R_\odot$ is the solar radius,
    and $L_\odot$ the solar luminosity).
%
%
    {\sffamily\textbf b,}~Period spacings between adjacent $l=1$ modes for
    the model marked with a vertical line in~{\bf a}.  The period spacing
    of the g-dominated modes (\DPg) can be seen from the maximum values to
    be about 75\,s.  Note that model frequencies were not corrected for
    near-surface effects\cite{KBChD2008,BBM2011}, which would have a small
    effect on the period spacings.
    {\sffamily\textbf c,}~\'Echelle diagram of $l=1$ modes for the same
    model as shown in~{\bf b}.  Here, the oscillation spectrum has been
    divided into segments of fixed length that are stacked one above the
    other.  Note that the abscissa is the period modulo the g-mode period
    spacing,~\DPg\ (whereas a conventional \'echelle diagram plots
    frequencies modulo the p-mode frequency spacing,~\Dnu).
    {\sffamily\textbf d,}~\'Echelle diagram of observed $l=1$ frequencies in the star
    KIC~\paul.  We conclude that the true g-mode spacing is $\DPg =
    77.1$\,s, whereas the average spacing of the observed
    modes\cite{BBM2011} was found to be $\DPobs \approx 55$\,s.
    \label{fig.model}}
\end{figure}

Figure~\ref{fig.model}a shows theoretical oscillation frequencies in a red
giant with a mass of 1.5$M_\odot$ (where $M_\odot$ is the solar mass).  The
dashed lines show the radial modes ($l=0$), whose frequencies decrease with
time as the envelope of the star expands.  These are pure p~modes and are
approximately equally spaced in frequency, with a separation of~\Dnu.  The
solid lines show the much denser spectrum of dipole modes ($l=1$).  The
g-dominated mixed modes appear as supward-sloping lines whose frequencies
increase with time as the stellar core contracts.  These modes are
approximately equally spaced in period.  The downward-sloping features that
run parallel to the $l=0$ modes are produced by mode bumping: the
p-dominated mixed modes, with frequencies decreasing with age, undergo
avoided crossings\cite{ASW77} with the g-dominated mixed modes.  (This
results in deviation from their mostly parallel appearance.)  A similar
pattern of mode bumping and avoided crossings is seen in models of subgiant
stars\cite{ChD2004,D+M2010}.



In Fig.~\ref{fig.model}b we show the period spacings between adjacent $l=1$
modes in one of the models, indicated in Fig.~\ref{fig.model}a with the
vertical line.  The dips in panel b correspond to bumped modes that are
squeezed together.  The period spacing of the g-dominated modes (\DPg) can
be measured from the upper envelope but cannot be observed directly because
only the bumped modes have enough p-mode character to be detected, by
virtue of their reduced mode
inertias\citep{DGH2001,ChD2004,DBS2009,BBM2011}.  Observations will detect
only a few modes in each p-mode order, and the average spacings of those
observable \new{sequences} (\DPobs) will be less than the true g-mode
spacing by up to a factor of two (the actual value depends on the number of
modes detected, \new{which is a function of the signal-to-noise in the
data,} and on the strength of the coupling between the g- and p-mode
cavities\cite{D+M2010}).  Figure~\ref{fig.model}c shows the mode
frequencies of the model in Fig.~\ref{fig.model}b displayed in \'echelle
format, where the spectrum has been divided into segments that are stacked
one above the other.  Note that the abscissa shows the period modulo \DPg,
whereas a conventional \'echelle diagram plots frequencies modulo the
p-mode frequency spacing,~\Dnu.

\new{Sequences} of $l=1$ modes with approximately constant period spacings
were first observed in the red giant KIC~6928997 (ref.~\citen{BBM2011}) and
we have found similar patterns in several hundred more stars.  The
observations were obtained with the \kepler\ satellite over the first 13
months of its mission and were sampled every 29.4\,min in the long-cadence
mode\citep{JCC2010}.  Figure~\ref{fig.model}d shows the period \'echelle
diagram for KIC~6928997 and allows us to estimate the spacing of the
g-dominated~modes to be $\DPg = 77.1$\,s, which is the value required to
produce a vertical alignment.  Remarkably, we have been able to estimate
\DPg\ despite the fact that g-dominated~modes are not observed (the average
spacing of the observed modes\cite{BBM2011} is $\DPobs \approx 55$\,s).

\begin{figure}
  \begin{center}
    \includegraphics[height=0.43\textwidth]{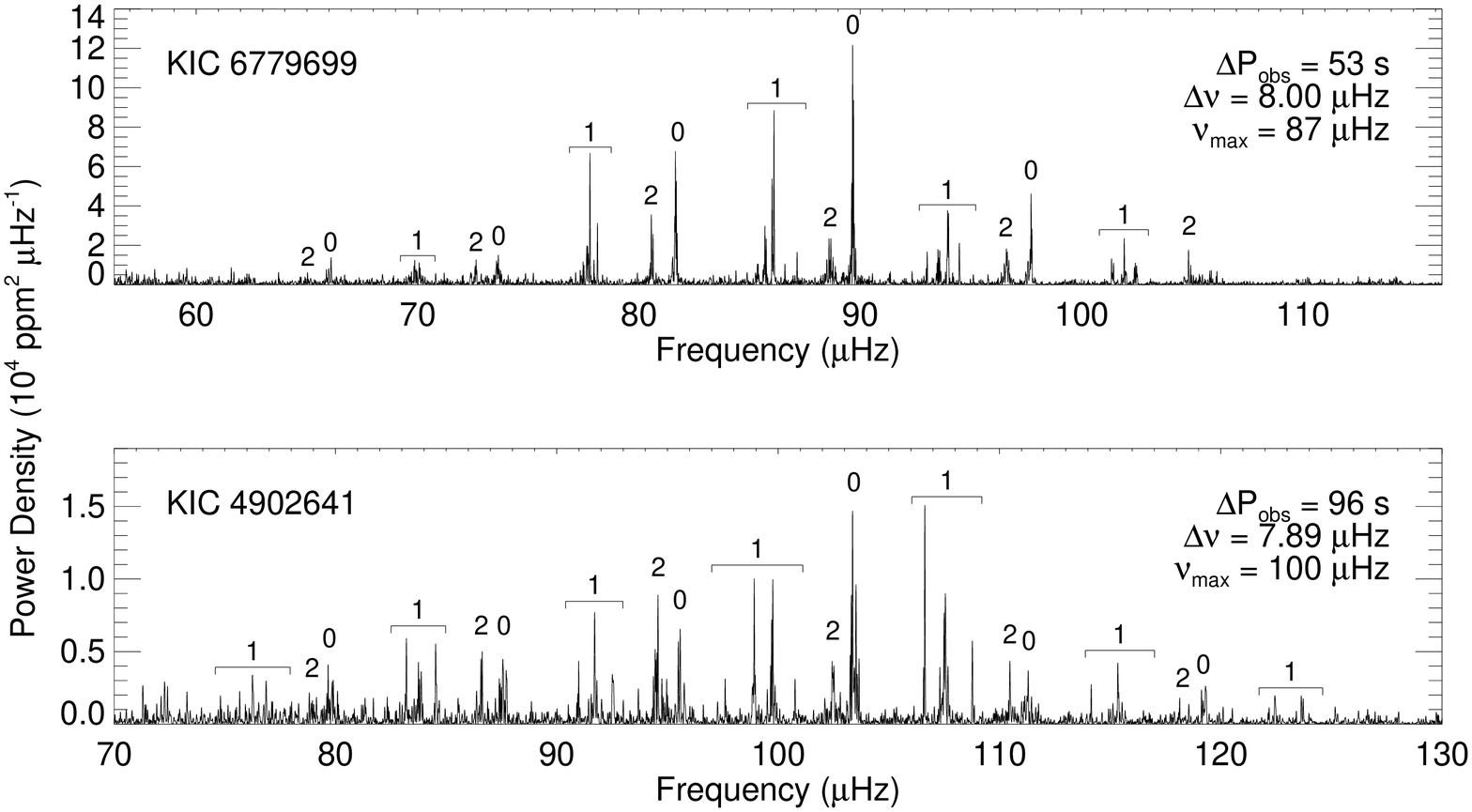}
    \includegraphics[height=0.43\textwidth]{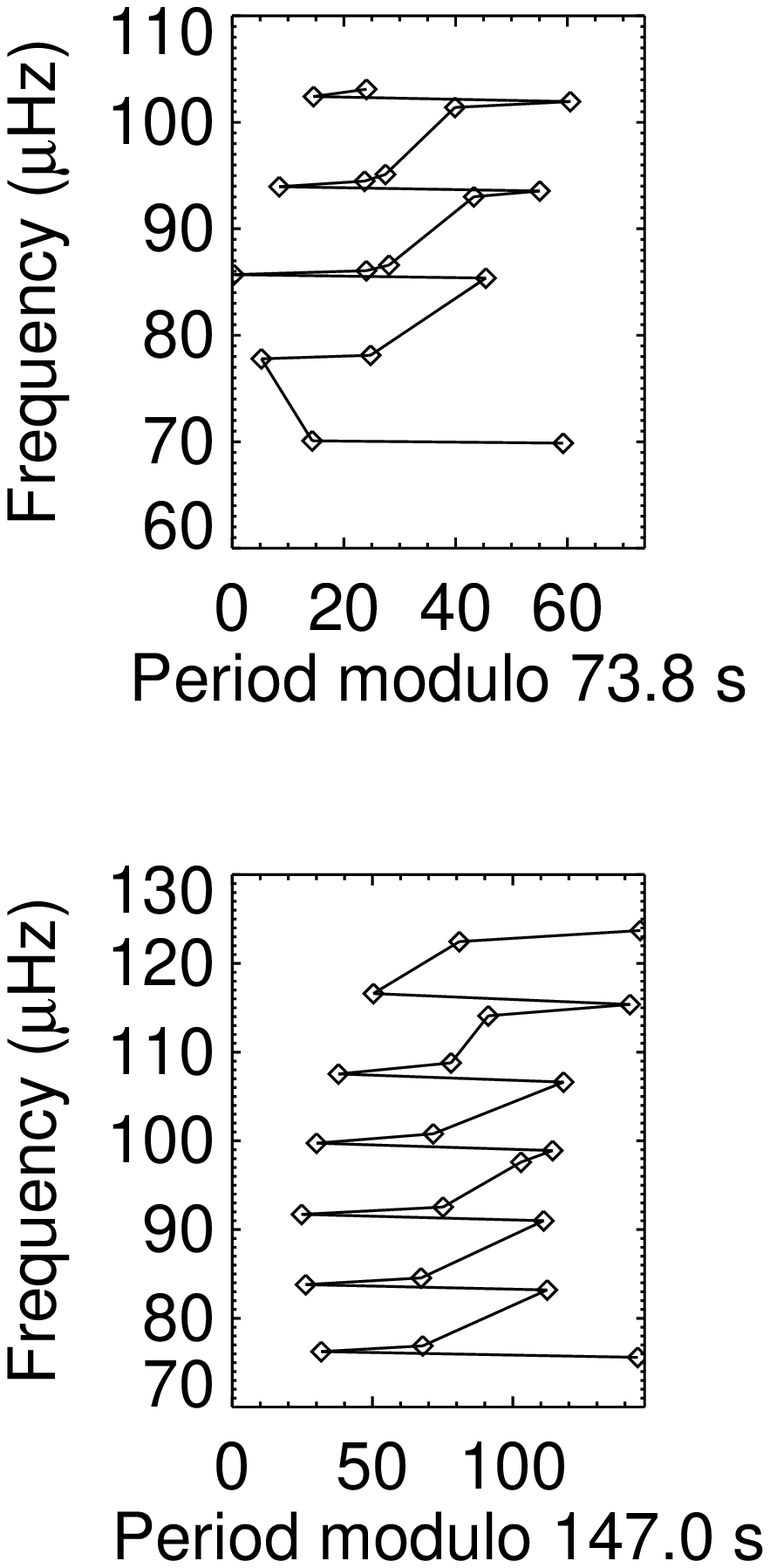}
  \end{center}

  \caption{{\sffamily\textbf{Oscillation power spectra and \'echelle
    diagrams of two red giant stars observed with \kepler.}}  Top,
    KIC~6779699; bottom, KIC~4902641: left, power spectra; right, échelle
    diagrams. The difference in the spacings of the $l=1$ modes indicates
    that KIC~6779699 is undergoing hydrogen-shell-burning on the red giant
    branch, while KIC~4902641 is \new{also} burning helium in its core (see
    Fig.~\ref{fig.results}).  Observations of KIC~6779699 were made over
    the first 13 months of the \kepler\ mission (Q0--Q5), while those of
    KIC~4902641 were made over the first 10 months (Q0--Q4).


%
    \label{fig.examples}}
\end{figure}

Figure~\ref{fig.examples} compares observed power spectra of two red giants
that have similar p-mode spacings ($\Dnu \approx 8$\,\muHz) but very
different $l=1$ period spacings.  We note that the outermost peaks in each
$l=1$ \new{cluster} (values of $l$ are given above the peaks), which we
expect to be the closest in character to the g-dominated modes, appear to
be the narrowest.  This observation is consistent with theoretical
calculations of mode inertias and
lifetimes\cite{ChD2004,DBS2009,MMN2010,DiMCC2011}.  Once again we have
detected enough modes to determine \DPg\ unambiguously using \'echelle
diagrams (right panels).  We find \DPg\ for the two stars to differ by
about a factor of two, implying they have very different core properties.

Inferring \DPg\ in this way using the period \'echelle diagram is not
possible for most of the stars in our sample, because it requires at least
3--4 modes to be detected in several of the $l=1$ \new{clusters, which is
only possible for the stars with the best signal-to-noise}.  Therefore, we
have instead measured the average period spacing of the observed $l=1$
modes (\DPobs) by using the power spectrum of the power spectrum.  In this
method, the power spectrum was first expressed in period rather than
frequency and then set to zero in regions not containing power from the
$l=1$ modes, as determined using the methods of ref.~\citen{HBS2010}.  The
power spectrum of this power spectrum was then calculated in order to
determine the most prominent period spacing.  For the reasons mentioned
above, we expect \DPobs\ to be less than \DPg\ and we have measured the
ratio between them to be in the range 1.3--1.6 in the few cases where \DPg\
can be estimated unambiguously.  Two other methods to measure \DPobs\ gave
comparable results.  The first was simply to measure pairwise separations
of the strongest $l=1$ peaks in the power spectrum.  The other was to
calculate the autocorrelation of the time series\cite{M+A2009} with narrow
filters centred on the $l=1$ \new{clusters}.

\begin{figure}
  \begin{center}
    \includegraphics[width=0.6\textwidth]{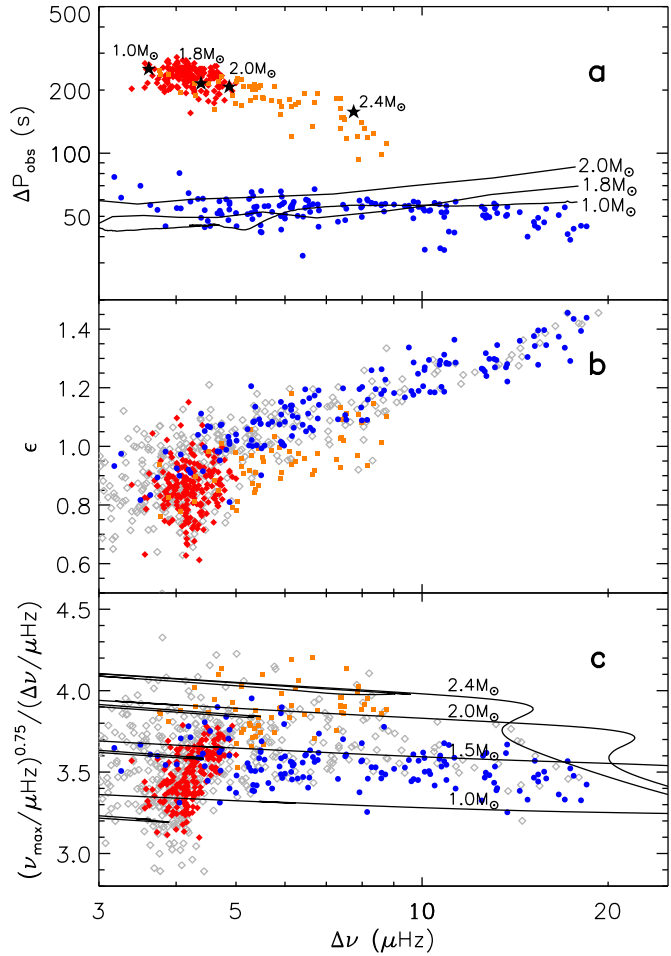}
  \end{center}
  
  \caption{{\sffamily\textbf{Asteroseismic diagrams for red giants observed with
    \kepler.}}  The abscissa is the p-mode large frequency separation.
    {\sffamily\textbf a,}~Filled symbols show period spacings measured from
    observations (a table listing the values is given in the Supplementary
    Information).  The stars divide into two clear groups, with blue
    circles indicating hydrogen-shell-burning giants (143 stars), while the
    remainder are helium-core-burning stars (193 red diamonds and 61 orange
    squares, divided on the basis of panel~c).  The solid lines show
    average observable period spacings for \textsc{astec}\cite{ChD2008a}
    models of hydrogen-shell-burning giants on the red giant branch as they
    evolve from right to left, calculated from the central three modes in
    the $l=1$ \new{clusters}.  The black stars show theoretical period
    spacings calculated in the same way, for four models of
    helium-core-burning stars that are midway through that phase (core
    helium fraction~50\%).  The 2.4-$M_\odot$ model was calculated with
    \textsc{astec}\cite{ChD2008a} and commenced helium-burning without
    passing through a helium flash.  The other three models, which did
    undergo a helium flash, were computed using the ATON
    code\cite{VDM2008,MMN2010} \new{(J.M. {\em et al.}, manuscript in
    preparation)}.  Solar metallicity was adopted for all models, which
    were computed without mass loss.
    {\sffamily\textbf b,}~The quantity $\epsilon$, which specifies the absolute
    frequency of the p-mode comb pattern\cite{HBS2010,MBG2011}.  We see a
    systematic difference between the hydrogen- and helium-burning stars.
    The open grey diamonds are stars for which a reliable period spacing
    could not be measured (391 stars).  Many of these had high
    signal-to-noise but lacked a clear structure that would indicate
    regular period spacings.
    {\sffamily\textbf c,}~The quantity $\numax^{0.75}/\Dnu$, which is
    approximately independent of luminosity\cite{HBS2010} (here \numax\ and
    \Dnu\ are in \muHz).  Helium-burning stars that we identify as
    belonging to the red clump, \new{based on their positions in this
    diagram,} are marked with red diamonds.  The remainder, which
    presumably belong to the secondary clump, are marked with orange
    squares.  The lines show model calculations based on scaling
    relations\cite{K+B95} for \Dnu\ and \numax\ applied to
    solar-metallicity BASTI models\cite{PCS2004} \new{with the mass loss
    efficiency parameter set to $\eta=0.2$}.

\label{fig.results}}
\end{figure}

Observed period spacings for about 400 stars are shown in
Fig.~\ref{fig.results}a, which clearly demonstrates the existence of two
distinct populations with different core properties.  A comparison with
model calculations confirms that the two groups coincide with
hydrogen-shell-burning stars on the red giant branch (blue circles) and
those that are also burning helium in the core (red diamonds and orange
squares).  We conclude that \DPobs\ is an extremely reliable parameter for
distinguishing between stars in these two evolutionary stages, which are
known to have very different core densities\cite{MMN2010} but are otherwise
very similar in their fundamental \new{properties} (mass, luminosity and
radius).  Note that other asteroseismic observables, such as the small
p-mode separations, are not able to do this\cite{HBS2010,MMN2010}.

Our ability to distinguish between hydrogen- and helium-burning stars makes
it possible to investigate their properties as separate populations.  One
example is the parameter $\epsilon$, which specifies the absolute position
of the p-mode comb pattern\cite{HBS2010,MBG2011}.  As shown in
Fig.~\ref{fig.results}b, there is a systematic offset between the two
populations.  This may indicate a difference in the surface layers, given
that $\epsilon$ is sensitive to the upper turning point of the
modes\cite{Gou86}.  \new{However, the difference may also arise because the
envelope of oscillation power is centred at different frequencies in the
two types of stars (see below).}  This result is clearly worthy of further
study.

A very important application for the helium-burning stars is to distinguish
between the so-called red clump and secondary clump\cite{Gir99,MMB2009}.
The red clump comprises low-mass stars that suffered from electron
degeneracy in their hydrogen-shell-burning phase and ignited helium in a
flash once the core attained a critical mass.  This common core mass
explains why the red clump (known as the horizontal branch when seen in
metal-poor clusters) spans a very narrow range of luminosities.  The
secondary-clump stars, meanwhile, are too massive to have undergone a
helium flash and so have a range of core masses, and hence of luminosities.
The mass threshold that divides these two populations depends on
metallicity, and also on core overshoot\cite{Gir99}.  

Among the helium-burning stars in Fig.~\ref{fig.results}a we can indeed see
this division into a compact group (the clump; red diamonds) and a
dispersed group (the secondary clump; orange squares).  It is even more
apparent when we examine the quantity $\numax^{0.75}/\Dnu$, which is
approximately independent of luminosity\cite{HBS2010} (\numax\ is the
frequency at which the oscillation envelope has its maximum\cite{K+B95}).
This quantity is shown in Fig.~\ref{fig.results}c and the comparison with
evolutionary models having solar metallicity implies a helium-flash
threshold of around $2\,M_\odot$.  Refinement of this result, using data
from more detailed studies of individual stars near the boundary between
the clump and secondary clump, should test predictions of convective-core
overshoot.

\small
{\sffamily

\begin{addendum}
\item[Supplementary Information] is linked to the online version of the
  paper at www.nature.com/nature.

\item[Acknowledgements] We acknowledge the entire \kepler\ team, whose
  efforts made these results possible.  We thank M. Biercuk for comments.
  Funding for this Discovery mission is provided by NASA's Science Mission
  Directorate.  T.R.B and D.S. were supported by the Australian Research
  Council; P.B. and C.A. were supported by the European Community's 7th
  Framework Programme (PROSPERITY); S.H. was supported by the Netherlands
  Organisation for Scientific Research (NWO).  The National Center for
  Atmospheric Research is sponsored by the National Science Foundation.

 \item[Author Contributions] 
T.R.B, B.M., P.B., Y.P.E, R.A.G., S.H., C.A., A.-M.B. and F.C. measured and
interpreted period spacings;
B.M., D.H., R.A.G., S.H., T.K., W.J.C., C.B., D.L.B. and S.M. calculated
power spectra and measured large frequency separations;
J.M., J.C.-D., A.M., D.S., T.R.W., K.B., M.P.D.M., M.-A.D., M.J.G., S.K.,
A.N., V.S.A. and P.V. calculated and interpreted theoretical models;
J.D.R., S.H., S.F., Y.P.E., D.S., T.M.B., H.K., J.C.-D. and R.L.G
contributed to the coordination of the project, including the acquisition
and distribution of the data;
J.M.J. constructed the photometric time series from the original
Kepler pixel data.
All authors discussed the results and commented on the manuscript.

\item[Author Information] Reprints and permissions information is available
  at www.nature.com/reprints. The authors declare that they have no
  competing financial interests. Correspondence and requests for materials
  should be addressed to T.R.B. (e-mail: t.bedding@physics.usyd.edu.au).
\end{addendum}
}

\newpage

\clearpage

\clearpage

\end{document}